# Grain Growth Kinetics in (Cr,Mo,Ta,V,W)C High-Entropy Carbide Ceramics

Ali Sarikhani[1,*], Gregory E. Hilmas[2], David W. Lipke[2], Douglas E. Wolfe[3], Stefano Curtarolo[4], Shen J. Dillon[5], Ahmad Mirzaei[5], William G. Fahrenholtz[2,*]

[1] *Missouri University of Science and Technology, Materials Research Center, Rolla, MO, USA*
[2] *Missouri University of Science and Technology, Materials Science and Engineering, Rolla, MO, USA*
[3] *Pennsylvania State University, Materials Science and Engineering, University Park, PA, USA*
[4] *Duke University, Mechanical Engineering and Materials Science, Center for Extreme Materials, Durham, NC, USA*
[5] *University of California, Irvine, Materials Science and Engineering, Irvine, CA, USA*

***ORCID:***

***Sarikhani**: 0000-0001-8121-0867*
***Hilmas**: 0000-0002-1611-5457*
***Lipke**: 0000-0002-4557-6690*
***Wolfe**: 0009-0004-5090-0290*
***Curtarolo**: 0000-0003-0570-8238*
***Dillon**: 0000-0002-6192-4026*
***Mirzaei**: 0000-0003-1518-9937*
***Fahrenholtz**: 0000-0002-8497-0092*

## Abstract

Understanding grain-boundary mobility during spark plasma sintering can enable microstructure control in high-entropy carbides, yet quantitative grain-growth kinetics remain scarce. In this work, grain growth kinetics and densification behavior are investigated for single-phase, near fully dense (Cr,Mo,Ta,V,W)C high-entropy carbide ceramics. Specimens were densified by spark plasma sintering for 10 min at 1750-1950 °C to isolate the effect of temperature on microstructural evolution. Increasing sintering temperature produced systematic grain growth, a small increase in lattice parameter, and reduced Ta segregation while maintaining a single-phase rock salt structure. Grain growth was analyzed using a normal grain growth model with an assumed exponent of $n$=3, yielding an apparent activation energy of approximately 620 kJ mol$^{-1}$. Densification curves showed rapid consolidation before reaching the peak temperature, followed by temperature-dominated grain coarsening. These results establish quantitative links between densification temperature, grain growth, and diffusion-assisted homogenization in high-entropy carbide ceramics.

**Keywords:** High-entropy carbide; Spark plasma sintering; Grain growth kinetics; Densification behavior; Apparent activation energy; Microstructural evolution.

* Corresponding authors: A. Sarikhani (as5kw@mst.edu), W. G. Fahrenholtz (billf@mst.edu)

## Introduction

High-entropy carbides (HECs) are carbide ultra-high temperature ceramics in which multiple transition metals occupy the metal sublattice of a single-phase carbide with the rock salt crystal structure [1-5]. These materials have attracted attention because they combine high melting temperatures, hardness, and chemical stability with the compositional flexibility of multi-principal-element design [1-4,6-9]. Building on concepts first established for high-entropy alloys [10-12], HECs are being explored for extreme-environment applications including aerospace thermal protection and other high-temperature structural uses [5,13].

Refractory carbide compositions containing group IV, V, and VI transition metals have shown strong potential for producing dense single-phase structures with robust mechanical performance [1,2,8,14-19]. In particular, (Cr,Mo,Ta,V,W)C is an attractive model system because it can be synthesized by carbothermal reduction and densified by spark plasma sintering (SPS) into a nominally single-phase ceramic with high hardness and stable thermal/electrical behavior [9,16-21]. Prior studies established processing routes and demonstrated that control of carbon-related defects/segregation can influence transport behavior [19,21]. However, those studies did not isolate the effect of sintering temperature on grain growth kinetics, densification progression, or chemical homogenization.

The composition used in this work was previously optimized to produce a single phase ceramic with no apparent residual oxide inclusions or excess carbon [20]. Prior processing optimization included several different carbon contents that were referenced to the nominally stoichiometric carbothermal reduction reaction [20], including the -7.5 wt.% carbon compositions used in the present study. The present study uses this optimized nominally carbon-deficient composition as a model system for evaluating SPS-temperature-driven grain growth, densification behavior, and chemical homogenization, not as a platform to independently isolate carbon vacancy content. Because carbothermal reduction can leave both carbon vacancies and/or oxygen on the C-site sublattice [22], the final material is more rigorously treated as a nominally carbon-deficient oxycarbide, (Cr,Mo,Ta,V,W)C [23,24]. The nominal composition described in the paper is based on the batching used for synthesis as determination of exact site occupancy and oxygen contents was beyond the scope of the present study.

Despite broad progress in synthesis and property characterization, quantitative grain-growth kinetics remain underreported for HECs [1,4,16-18,23-27]. This is an important research opportunity because microstructural evolution during sintering directly affects grain boundary density, defect equilibration, and local compositional heterogeneity, which, in turn, can influence functional and mechanical performance. In conventional ceramics, grain growth is often modeled as a diffusion-controlled process with a thermally activated growth factor; however, in multicomponent carbides, boundary mobility can be modified by solute drag, carbon-vacancy effects, oxygen incorporation, and local segregation [4,16-18,23-25,27,28]. As a result, apparent activation energies and grain growth exponents may differ substantially from simpler carbide systems containing just one transition metal. This need is particularly relevant for SPS-processed materials because SPS combines rapid heating, short dwell times, and applied pressure, enabling near-full densification over a narrow processing window while preserving measurable differences in grain size [13,16,17,23-26]. This allows densification and grain coarsening to be partially decoupled, making the final holding temperature a useful variable for comparative analysis of grain-growth behavior, while recognizing that quantitative kinetic interpretation still depends on assumptions about the evolving grain-size distribution and growth regime [23,24,26]. In addition, SPS processing histories can be analyzed using ram displacement data, providing insight into the densification regime and its transition to grain-growth-dominated behavior during the high-temperature dwell [23,24,26].

Accordingly, a controlled SPS temperature series with a fixed dwell time provides a direct route to quantifying grain growth kinetics in a chemically complex carbide while minimizing confounding effects from porosity or phase changes. Building on this framework, near fully dense (Cr,Mo,Ta,V,W)C ceramics were synthesized from a fixed batch composition and densified by SPS at temperatures between 1750 and 1950 °C. This work isolates sintering temperature as the primary processing variable affecting grain growth in a high-entropy carbide system.

## Experimental methods

### *Powder Processing and Carbothermal Reduction*

The processing route used in this work involved carbothermal reduction of mixed metal oxides with carbon to form the high-entropy carbide powder, followed by SPS densification of the reacted powder. This distinction is important because carbothermal reaction sintering can introduce chemical driving forces for microstructural evolution in addition to the surface-energy-driven effects typically considered in conventional sintering. The commercial oxide powders that were used as starting reagents were: chromium(III) oxide ($Cr_2O_3$, 99.5%, 0.7 µm; Elementis, Corpus Christi, TX); molybdenum(VI) oxide ($MoO_3$, 99.9%, 6 µm; US Research Nanomaterials, Houston, TX); tantalum(V) oxide ($Ta_2O_5$, 99.8%, 1-5 µm; Atlantic Equipment Engineers, Upper Saddle River, NJ); vanadium(V) oxide ($V_2O_5$, 99.6%, -10-mesh; Alfa Aesar); and tungsten(VI) oxide ($WO_3$, 99.9%, ~80 nm; Inframat Advanced Materials, Manchester, CT). Carbon black (C120; Cabot, Alpharetta, GA) served as the carbon source and reductant. The powders were blended by high-energy ball milling (SPEX, 8000D MIXER/MILL, Glen Mills Inc.) in a WC vial using WC milling media, with a WC media-to-powder mass ratio of 15:2. Milling was performed as two 1 h runs separated by a 0.5 h cooling interval. Two identical 20 g batches were prepared. After milling, the powder was passed through a 60-mesh sieve and then compacted into pellets using steel die sets. The pellets were reacted simultaneously at 1610 °C for 3 h under a mild vacuum of ~13.3 Pa in a graphite-element sintering furnace (HP50-7010 G, Thermal Technology).

The carbothermal reactions followed the generalized reaction scheme shown in Eq. 1. The compositions were optimized with nominally carbon-deficient batches to produce single-phase ceramics with no residual oxides and no excess carbon inclusions:

$$0.5\,Cr_2O_3 + MoO_3 + 0.5\,Ta_2O_5 + 0.5\,V_2O_5 + WO_3 + (5x - 5y + 12.5)\,C \rightarrow$$

$$5\,(Cr_{0.2}Mo_{0.2}Ta_{0.2}V_{0.2}W_{0.2})C_xO_y + (12.5 - 5y)\,CO \qquad (1).$$

The 7.5 wt.% carbon-deficient batching condition was selected based on prior processing optimization of this composition because it produced dense, nominally single-phase rock salt carbide ceramics without obvious residual oxide or free-carbon inclusions [20]. Thus, the carbon deficiency used here is a processing/batching parameter rather than a directly measured carbon-vacancy concentration. In addition, synthesis by carbothermal reduction of mixed oxide precursors results in residual oxygen in the reacted powder [22]. Therefore, the final composition is described in this work as a nominally carbon-deficient high-entropy oxycarbide, but is referred to as (Cr,Mo,Ta,V,W)C herein for brevity. The reduced carbon content relative to the stoichiometric amount was sufficient for complete carbothermal reduction and carbide formation. If the C-site sublattice occupancy were assumed from the nominal batch composition and the carbon shortfall were accommodated entirely by oxygen incorporation, the nominal batching calculation would correspond approximately to (Cr,Mo,Ta,V,W)$C_{0.87}O_{0.13}$. If oxygen incorporation were neglected, the same batching condition would instead correspond to a larger nominal carbon-vacancy concentration. The final composition is also impacted by the SPS process, which is conducted in graphite dies (i.e., a nominal carbon activity of one). This could decrease oxygen content and increase carbon content by reaction during the densification cycle. Because the actual carbon and oxygen contents were not independently quantified in the present work, the values of x, y, and the carbon-vacancy concentration were not assigned quantitatively and the resulting ceramic is referred to as (Cr,Mo,Ta,V,W)C to be consistent with the typical way that similar materials are presented in previously published papers [16,19,29].

*Spark Plasma Sintering and Densification Analysis*

The reacted pellets were crushed, passed through a 100-mesh sieve, divided into five portions, and densified by SPS at five temperatures ranging from 1750 to 1950 °C. The SPS cycle began under vacuum (<6 Pa) with heating to 1600 °C at 100 °C/min under an initial uniaxial pressure of 15 MPa, followed by a 5 min dwell. The pressure was then increased to 50 MPa at 35 MPa/min, and the temperature was raised to the final sintering temperature (1750-1950 °C) at 100 °C/min under constant pressure. After a 10 min dwell at the peak temperature, the pressure was reduced to 25 MPa at 25 MPa/min while cooling to 1200 °C at 50 °C/min. Ram displacement was recorded throughout the SPS cycle and used to calculate densification curves. For SPS densification analysis, the ram displacement data were corrected for thermal expansion of the tooling assembly. The thermal expansion correction factor was determined from the slope of ram position versus pyrometer temperature during the 25 MPa stage, using a linear fit of the selected region. The corrected displacement at each temperature was calculated from the measured ram displacement and the thermal expansion contribution of the die setup. The net sample shrinkage, $\Delta L$, was then obtained from the difference between the corrected displacement at the end of sintering and that at a given time. Assuming uniaxial shrinkage along the pressing direction, the instantaneous relative density was estimated from the final density and the normalized change in sample thickness, using an initial constant thickness. This procedure was used to generate the densification curve as a function of temperature.

*Grain Growth Analysis*

The grain growth kinetics were analyzed using the measured grain sizes for each specimen. The grain growth factor, $K$, was extracted from the empirical relation:

$$G^n = G_0^n + Kt \;\; [\mu\text{m}^n] \qquad (2),$$

where $G$ is the average grain size, $G_0$ is the initial mean grain size at the chosen $t$=0 for the grain-growth analysis, and $t$ is the isothermal sintering time. In this framework, $t$=0 is an analysis reference state rather than the literal start of sintering, and its use assumes that the material has already reached a dense microstructure and entered the relevant grain-growth regime. Because the initial mean grain size corresponding to this reference state ($G_0$) was not independently measured, the present analysis assumes $G^n \gg G_0^n$ and neglects the $G_0^n$ term, a common approximation in grain-growth analyses [30]. Because the initial grain-size distribution at the chosen $t$=0 could not be measured directly, the present analysis does not independently verify steady-state, self-similar grain growth and should therefore be interpreted as an apparent grain-growth analysis based on an assumed growth law. The exponent $n$ describes the isothermal time dependence of grain growth within the assumed growth law; however, relating a particular $n$ value to a specific grain-growth mechanism requires additional assumptions regarding how the microstructure evolves, including whether the system is in a steady-state regime and whether effects such as pore drag, solute drag, or defect pinning influence boundary mobility. Finally, $K$ is the kinetic growth factor, which is assumed to follow an Arrhenius-type temperature dependence:

$$K = K_0 \, e^{-E/RT} \left[\frac{\mu m^n}{min}\right] \quad (3),$$

in which, $E$ is the apparent activation energy for grain boundary migration, $T$ is the sintering temperature, and $R$ is the universal gas constant. These equations represent an assumed normal-grain-growth framework, in which the grain-size distribution evolves in an approximately self-similar manner under topological constraints.

### *Microstructural Characterization*

High-contrast images of the microstructures used for grain size analysis were obtained by scanning electron microscopy (SEM; Axia ChemiSEM, Thermo Fisher Scientific Inc.). Grain boundaries were traced using image manipulation software (GIMP, The GNU Image Manipulation Program Team) and the Feret diameters were determined using image analysis software (ImageJ, National Institutes of Health, Bethesda, MD). The Feret diameter and standard error of the mean were determined by analyzing 200 to 600 grains for each sintering condition. Additional SEM images and chemical analysis using energy dispersive spectroscopy were performed on a separate instrument (PIONEER, Raith 150 eLine Plus).

*Phase and Density Characterization*

Phase analysis and lattice-parameter determination were performed by room-temperature X-ray diffraction (XRD; X'Pert MPD, Philips). Lattice parameters were analyzed by crushing sintered pellets and passing the resulting powder through a 100-mesh sieve. Powders were mixed with alumina as an internal standard ($Al_2O_3$, 99.8%, D50 ~0.6 µm, alpha alumina; Almatis, Leetsdale, PA). Rietveld refinement was used to determine lattice parameters with error determined from the known lattice parameters of the standard. The theoretical density was determined from the lattice parameter and the EDS-derived average metal-site composition assuming a rock salt carbide structure. The uncertainty in the theoretical density was estimated from the standard error of the mean of the EDS-derived metal-site atomic percentages collected from multiple measurements for each sample, with that compositional uncertainty propagated through the theoretical-density calculation. Bulk density was measured by the Archimedes method, and standard error of the mean was obtained by propagating the statistical uncertainty associated with each mass measurement used in the calculation. Relative density was then calculated from the ratio of bulk density to theoretical density, and its uncertainty was estimated by propagating the standard error of the mean of the bulk and theoretical densities. Based on the measured mass change of the WC milling media before and after milling, each specimen incorporated an average of ~0.03 g WC contamination; this contribution was accounted for in the corrected theoretical density and relative density calculations. Specimen nomenclature, average grain size (mean Feret diameter), and density values are summarized in Table 1.

| Sample | Sintering temperature (°C) | Avg. grain size (μm) | Theoretical density ($g/cm^3$) | Bulk density ($g/cm^3$) | Relative density (%) | Open porosity (%) |
|---|---|---|---|---|---|---|
| **S1750** | 1750 | 9.3 ± 0.3 | 10.5 ± 0.3 | 10.51 ± 0.03 | 100 ± 3 | 0.12 |
| **S1800** | 1800 | 13.4 ± 0.5 | 10.4 ± 0.2 | 10.46 ± 0.04 | 100 ± 2 | 0.38 |
| **S1850** | 1850 | 18.5 ± 0.5 | 10.3 ± 0.2 | 10.53 ± 0.17 | 102 ± 3 | 1.42 |
| **S1900** | 1900 | 21.5 ± 0.7 | 10.7 ± 0.3 | 10.66 ± 0.03 | 100 ± 3 | 0.16 |
| **S1950** | 1950 | 28.8 ± 0.7 | 10.8 ± 0.2 | 10.55 ± 0.01 | 98 ± 2 | 0.11 |

**Table 1.** Sintering temperature, average grain size, WC-corrected theoretical density, bulk density, relative density, and measured open porosity of (Cr,Mo,Ta,V,W)C ceramics used for grain-growth kinetic analysis.

## Results and discussion

*Microstructure, SPS Densification Behavior, and Final Density*

Near fully dense, nominally carbon-deficient (Cr,Mo,Ta,V,W)C ceramics were obtained for each densification temperature (1750-1950 °C) using the same 10 min dwell time. The measured open porosity values were low for all specimens ranged from 0.11% to 1.42%, which confirmed that the SPS conditions produced near fully dense ceramics. The theoretical densities reported in Table 1 were corrected for the measured WC contamination introduced from the milling media. Relative density values close to or slightly above 100% are therefore interpreted as full density within the combined uncertainty of the Archimedes bulk-density measurement, lattice-parameter refinement, EDS-derived composition, and WC-contamination correction. Because all samples reached essentially the same final density, differences in microstructure and chemical homogeneity were interpreted as temperature-driven effects rather than porosity-driven effects. Likewise, the absence of significant quantities of residual oxides or carbon-rich inclusions indicates that results should be representative of intrinsic behavior of this composition, not extrinsic effects such as grain pinning from inclusions.

Back-scattered SEM imaging confirmed dense microstructures, consistent with the low open porosity values reported in Table 1. The SEM images also revealed an increase in grain size with increasing SPS temperature (Fig. 1). The average grain sizes increased from 9.3 ± 0.3 μm after SPS at 1750°C to 28.8 ± 0.7 μm after SPS at 1950 °C, consistent with thermally activated grain-boundary migration under fixed dwell conditions.

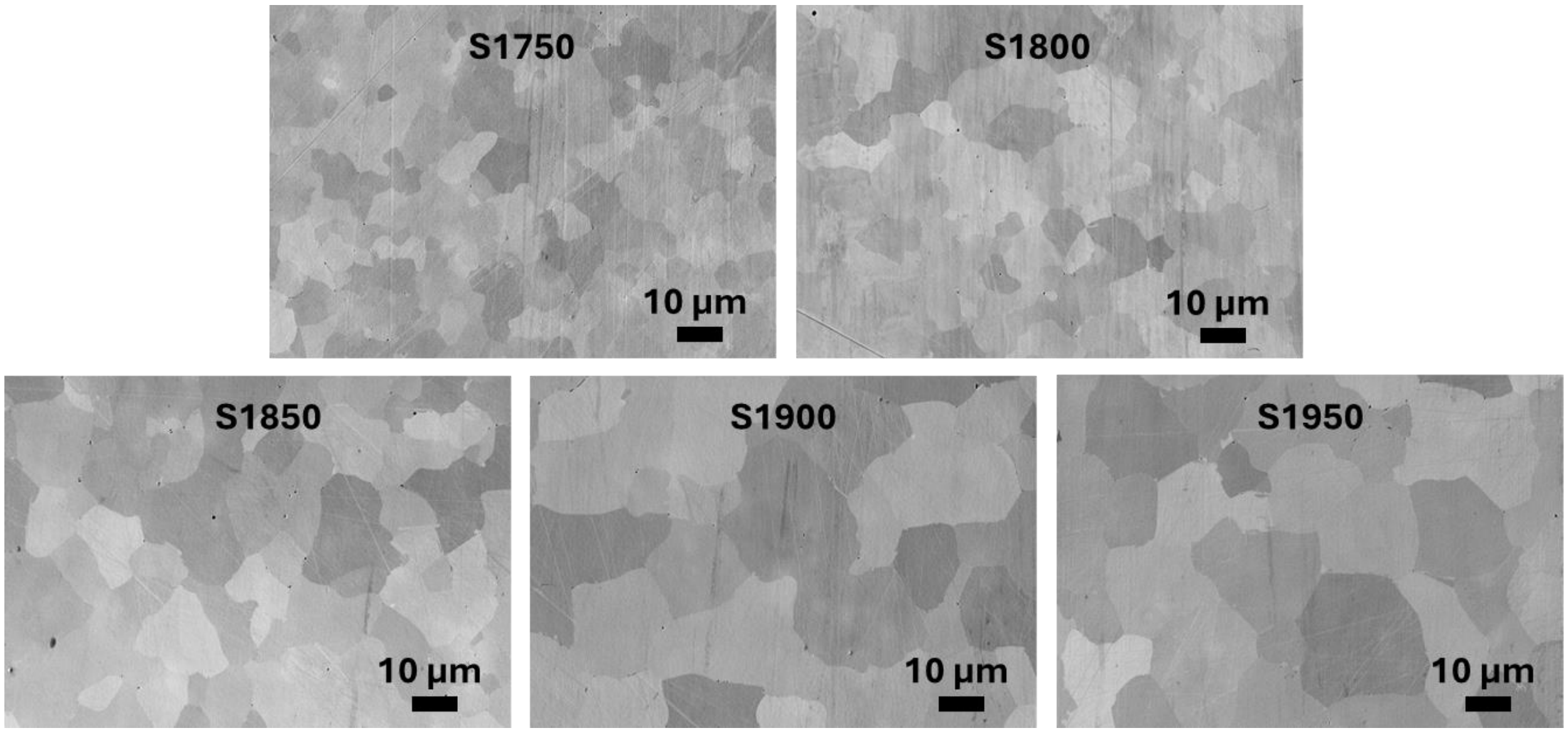

**Figure 1.** Microstructural evolution of (Cr,Mo,Ta,V,W)C with SPS temperature. Back-scattered SEM micrographs of specimens sintered at 1750, 1800, 1850, 1900, and 1950 °C for a constant 10 min dwell, showing near fully dense microstructures and monotonic grain coarsening with increasing densification temperature.

*Crystal Structure, Phase Stability, and Lattice Parameter Evolution*

Room-temperature XRD patterns from all specimens were indexed to a single-phase rock salt FCC structure, confirming phase stability of (Cr,Mo,Ta,V,W)C over the investigated temperature range (Fig. 2a). The diffraction profiles exhibited the characteristic rock salt selection rule (all-odd or all-even Miller indices of the diffracting planes) with the major peaks corresponding to the (111), (200), (220), (311), and (222) planes, and no detectable peak splitting, consistent with a single cubic phase across all sintering temperatures. No additional crystalline phases were detected within the XRD detection limit, indicating that the temperature sweep primarily altered microstructural features and did not change the phase constitution.

Rietveld refinement showed a small but systematic increase in lattice parameter with increasing sintering temperature ranging from 4.2767 Å for S1750 to 4.2806 Å for S1950 (Fig. 2b). To make this small lattice expansion more visible, the inset in Fig. 2b shows a magnified view of the rock salt FCC peak near $2\theta \approx 73$ °, after aligning the nearby $\alpha$-$Al_2O_3$ internal-standard peak at $2\theta \approx 67$ °, highlighting the systematic peak shift with increasing SPS temperature. Because all specimens retained the same rock salt phase and were prepared from the same nominal composition, the lattice expansion is not attributed to a phase transformation. Instead, it is interpreted as a correlated structural change associated with high-temperature SPS processing. Possible contributors include temperature-assisted chemical homogenization, reduction of local compositional partitioning, relaxation of microstrain introduced during powder synthesis and densification, and changes in the average local bonding environment of the multicomponent carbide lattice. Possible changes in the C-site sublattice occupancy involving carbon vacancies and/or oxygen incorporation may also contribute to the refined lattice-parameter trend. However, because the present study did not independently measure carbon and oxygen contents, the individual contributions of these effects cannot be separated. Therefore, the lattice expansion is discussed as a processing-correlated structural trend rather than direct evidence for a specific defect mechanism.

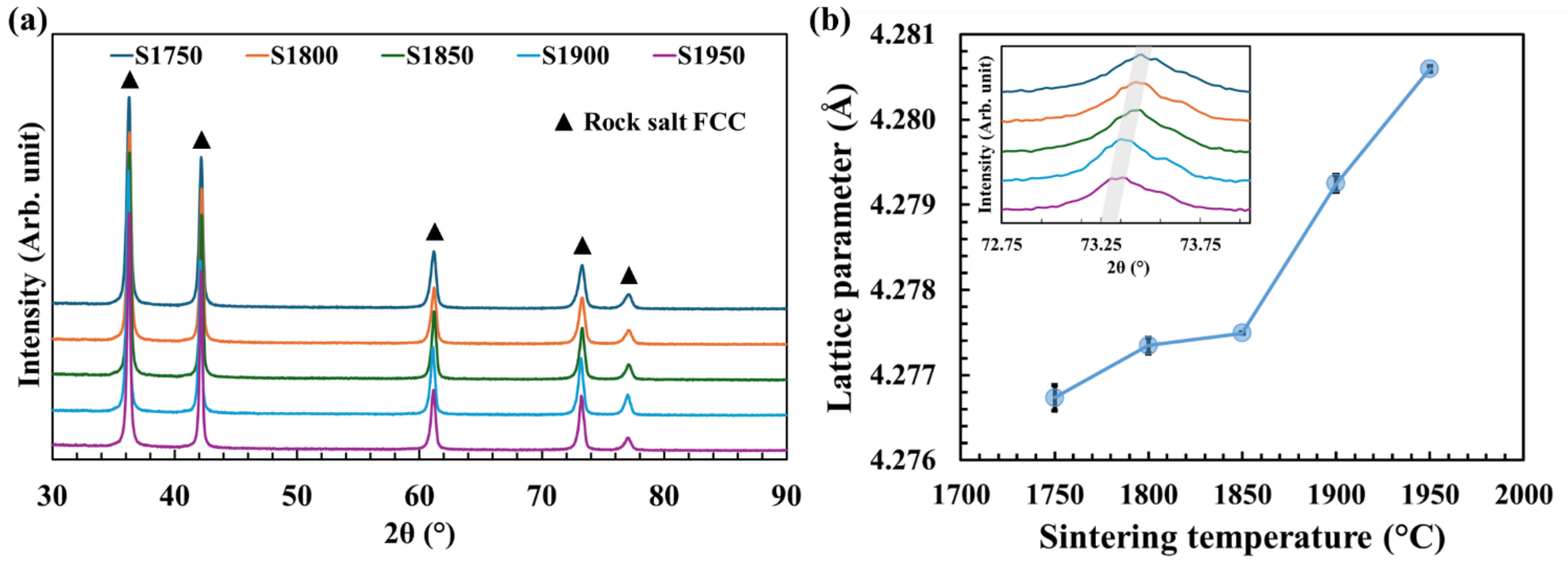

**Figure 2.** Crystal structure and lattice parameter evolution of (Cr,Mo,Ta,V,W)C with SPS temperature. (a) Room-temperature XRD patterns for specimens sintered at 1750-1950 °C, indexed to a single-phase rock salt (NaCl-type) FCC structure with major peaks corresponding to (111), (200), (220), (311), and (222). (b) Lattice parameters obtained by Rietveld refinement using α-$Al_2O_3$ as an internal standard, showing a systematic increase with densification temperature. The inset in (b) shows a magnified view of the rock salt FCC peak near 2θ ≈ 73 °, after alignment using the α-$Al_2O_3$ internal-standard peak near 2θ ≈ 67 °.

*Elemental Mapping and Chemical Homogenization*

Fig. 3 shows a representative micrograph of S1950 together with EDS elemental maps, indicating nominally uniform distributions of the metal species across the analyzed field of view. In addition to the maps, Fig. 3 reports the average EDS atomic percentages obtained from multiple measurements/locations within the specimen; the plotted values therefore represent bulk-averaged composition rather than a single spot analysis. The average EDS compositions were included to verify that the specimens remained broadly comparable in metal-site composition across the SPS temperature series, since all samples were prepared from the same starting batch. The measured compositions were generally similar among specimens, indicating that large changes in the average metal ratio did not occur during sintering. However, samples sintered at higher temperatures tended to show lower Cr contents in the EDS quantification, which may be associated with partial Cr volatilization during high-temperature processing. The average EDS compositions also consistently showed a higher apparent V content than the other transition metals. Because this trend was present for all specimens, all materials originated from the same powder batch, and neither the XRD results nor the EDS maps indicated V-rich secondary phases or localized V segregation, the higher apparent V content is interpreted as a systematic characteristic of the SEM/EDS quantification and/or a small deviation in the average bulk composition rather than a temperature-dependent segregation phenomenon. Thus, the average EDS data provide a check on compositional consistency while also identifying minor temperature-dependent composition changes that may accompany high-temperature SPS processing. The associated scatter (shown as statistical variation from the repeated measurements) reflects the combined effects of local microstructural heterogeneity, counting statistics, and detector/interaction-volume uncertainty, and provides a quantitative estimate of the degree of compositional uniformity at the scale accessible by SEM/EDS. Consistent with high-temperature equilibration, Fig. 3 shows that S1950 approaches a stable, uniform composition across the scanned regions. The full set of same-magnification SEM images and EDS elemental maps for Cr, Mo, Ta, V, W, C, and O in all specimens is provided in Supplementary Fig. S1 to enable direct comparison of elemental distributions across the SPS temperature series.

While the EDS maps in Fig. 3 indicate nominally uniform elemental distributions over the mapped field of view, the EDS map contrast and repeated local EDS measurements suggested that a small residual compositional inhomogeneity remained, most evidently in Ta. Thus, although Fig. 3 supports the conclusion that the material is macroscopically homogenized and single phase, it does not by itself resolve the extent of localized enrichment and depletion. Because Ta produces the strongest compositional contrast in the present system, it provides the clearest marker for assessing this residual partitioning. For that reason, the analysis in Fig. 4 focuses specifically on Ta-rich and Ta-deficient regions to quantify how the magnitude of local partitioning evolves with SPS temperature.

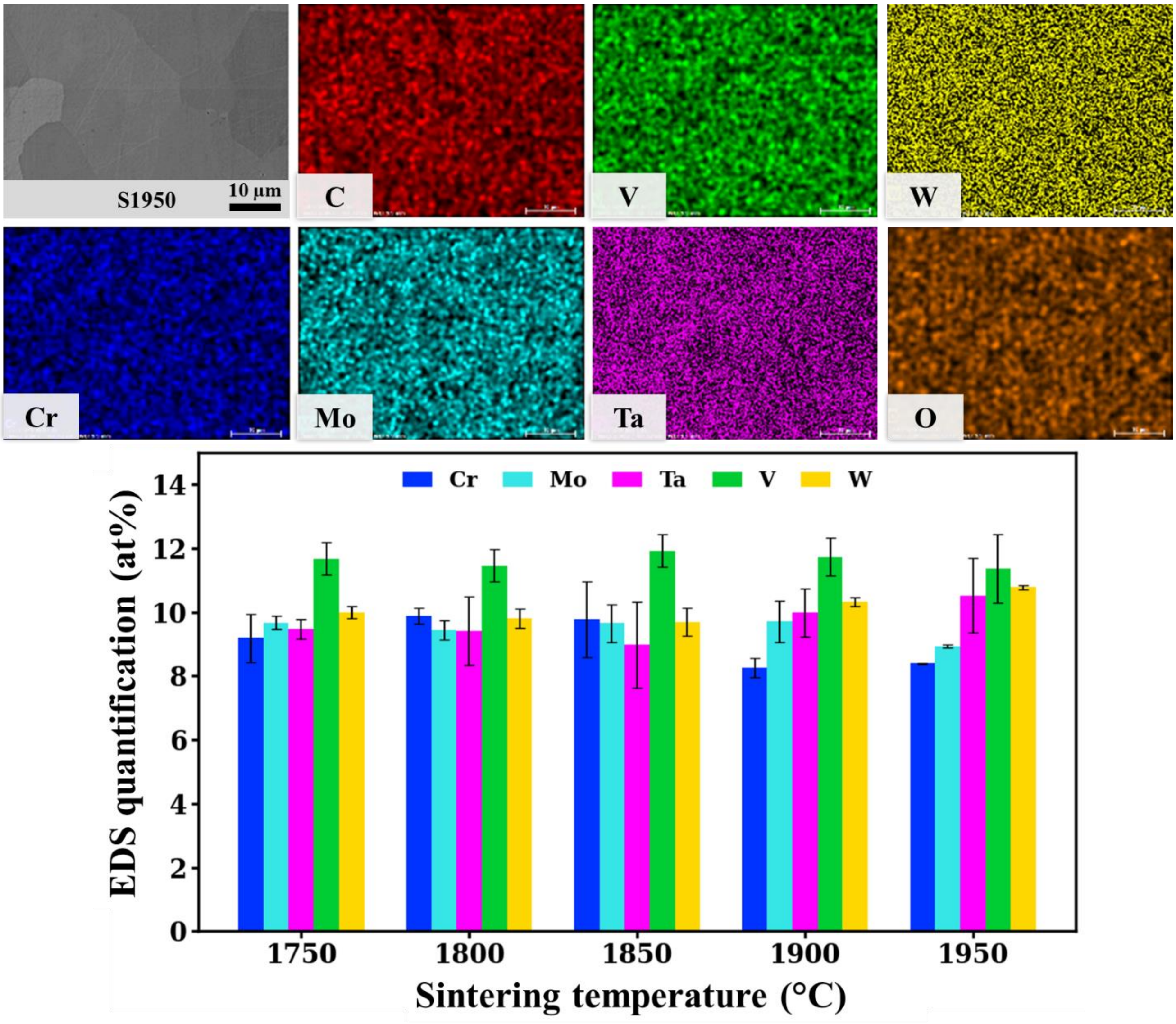

**Figure 3.** Elemental homogeneity and average EDS quantification for the S1950 specimen. Representative EDS elemental maps for the specimen sintered at 1950 °C (S1950), showing spatially uniform distributions of Cr, Mo,

Ta, V, and W across the analyzed field of view. The accompanying plot reports the average elemental composition measured by EDS (all-atom basis) for all specimens. Error bars represent the standard deviation from multiple EDS measurements.

To quantify the magnitude of local Ta partitioning, Fig. 4 summarizes the EDS point-scan quantification by separating measurements taken from visually identified Ta-rich regions and Ta-deficient regions, and then plotting the temperature dependence of each population along with their difference (shown as Ta content difference in this figure). Notably, the Ta-rich values at lower temperatures (S1750) approach ~12.9 at% (all-atom basis), whereas Ta-deficient regions are substantially lower, i.e., ~9.6 at%. For the specimen sintered at 1950 °C, the Ta-rich and Ta-deficient populations converge toward a narrower composition range (with a Ta content difference of ~1.5 at%), indicating improved uniformity. The decreasing Ta content difference with increasing SPS temperature indicated that the amplitude of local Ta partitioning was reduced, consistent with diffusion-assisted homogenization during high-temperature exposure. This trend is qualitatively consistent with the grain growth kinetics analysis, since enhanced short-range diffusion at higher temperatures could promote both grain-boundary migration and chemical redistribution; however, the present data do not independently rule out temperature-dependent grain-boundary segregation effects.

The combined XRD and EDS results therefore indicate that the primary effects of increasing SPS temperature were microstructural coarsening and chemical equilibration within a stable single-phase HEC matrix. No phase transitions or changes in relative density were observed, which helps define the scientific identity of the present work: the kinetics of microstructural evolution, not compositional optimization.

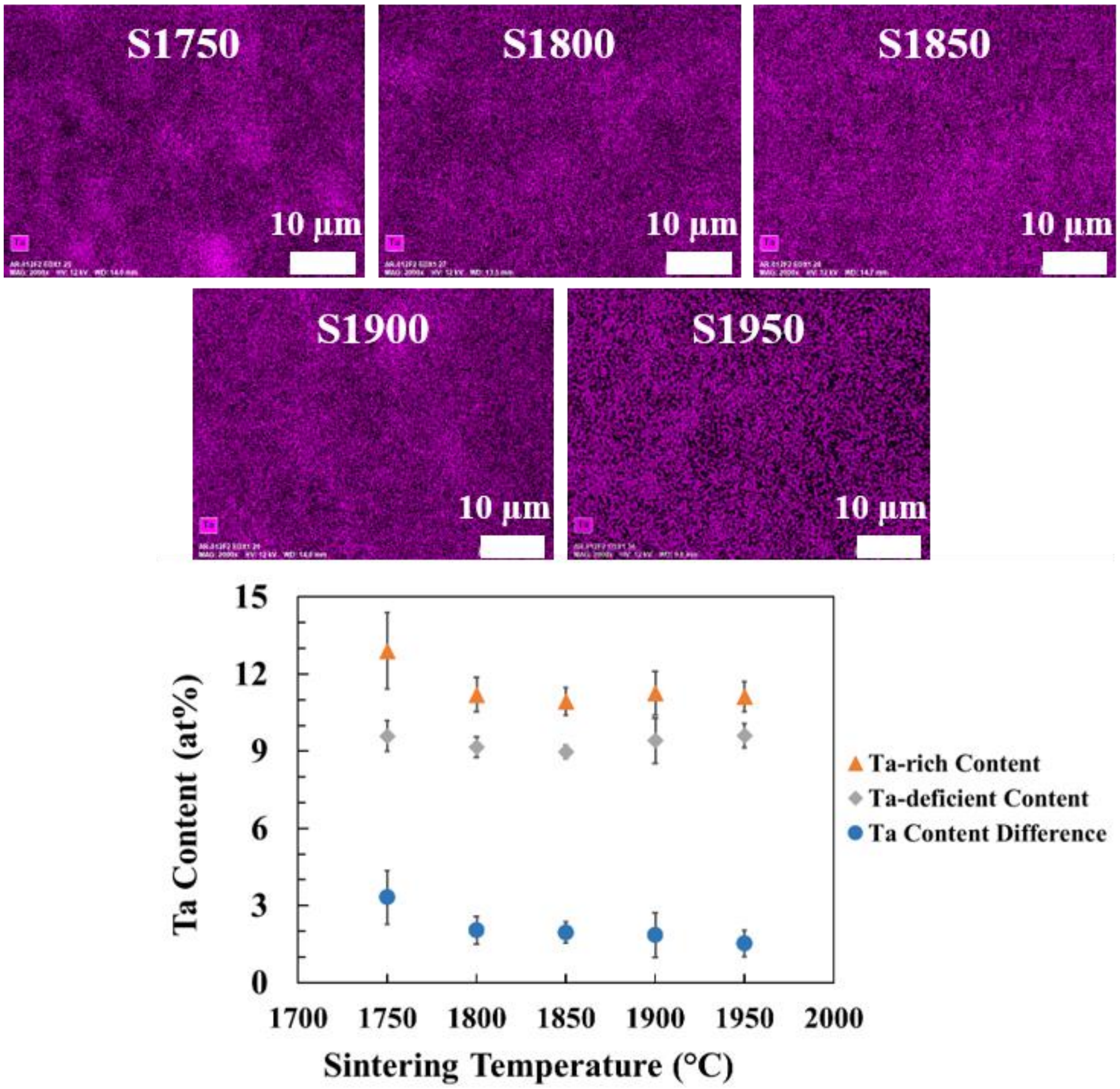

**Figure 4.** Temperature-dependent Ta partitioning quantified by EDS point analysis. Same-magnification Ta EDS maps are shown for specimens sintered at 1750, 1800, 1850, 1900, and 1950 °C to enable direct visual comparison of local Ta distributions. The plot summarizes Ta contents measured in visually identified Ta-rich and Ta-deficient regions, along with the Ta content difference defined as $\Delta$Ta = Ta-rich - Ta-deficient. The decreasing Ta content difference with increasing SPS temperature indicates reduced local Ta partitioning and improved chemical homogenization at higher sintering temperatures.

*Grain Growth Kinetics and Apparent Activation Energy*

The SPS ram-displacement-derived densification curves (Fig. 5a) show that most densification occurred prior to or near the final set temperature, while the subsequent hold at the densification temperature contributed relatively little shrinkage. Because densification was largely completed before the end of the dwell, the 10 min hold at peak temperature primarily promoted grain growth and local chemical equilibration, consistent with prior SPS densification analyses and constitutive modeling studies [31,32]. This partial decoupling of densification and coarsening enables grain growth to be evaluated under conditions where porosity is no longer the dominant variable [29,30]. Similar separation between densification and later coarsening/equilibration has been reported for other carbide and oxycarbide systems [23,24,26], supporting interpretation of the present temperature series as primarily microstructure-controlled rather than densification-limited.

To further assess the applicability of the assumed normal-grain-growth framework, the measured grain-size distributions were compared after normalization by their respective mean Feret diameters (Fig. 5b). The normalized cumulative distributions were broadly similar in shape across the SPS temperature series, indicating that the measured grain-size populations evolved in an approximately self-similar manner over the analyzed conditions. This observation provides indirect support for applying a common grain-growth law to the present dataset. However, the overlap was not exact, and the initial grain-size distribution at the chosen $t$=0 for the growth analysis could not be measured directly. Accordingly, the present treatment should be interpreted as an apparent grain-growth analysis based on an assumed normal-grain-growth framework, rather than a definitive verification of steady-state, self-similar grain growth.

The different final SPS temperatures provide a direct basis for evaluating grain growth kinetics as the specimens reached nearly full density before the dwell and the dwell time was the same for all specimens. A linear fit of *ln K* versus *1/T* (Fig. 5c) yielded an apparent activation energy of approximately 620 ± 40 kJ $mol^{-1}$, based on a growth exponent of $n$=3. The linear fit yielded a high coefficient of determination, $R^2$=0.98, consistent with an Arrhenius-like trend over the measured temperature range; however, the small number of data points limits the significance of the fit quality. The grain growth, combined with Arrhenius-type behavior, is consistent with thermally activated grain-boundary migration, but the fixed-time SPS dataset does not independently identify the specific rate-limiting diffusion pathway [23-27].

The grain growth exponent was assumed to be $n$=3, and the resulting kinetic parameters were interpreted within an apparent normal-grain-growth framework without direct proof of a specific atomic mechanism. In ceramics, a growth exponent of $n$=2 is commonly associated with ideal parabolic grain growth, in which curvature-driven grain-boundary migration occurs with approximately constant boundary mobility and without substantial drag or pinning effects. Deviations from this behavior can arise from several effects, including solute drag, local chemical partitioning, residual porosity, evolving grain-size distributions, or other mobility-limiting processes. In the present HEC system, the effective grain-boundary mobility may be influenced by metal-site chemical complexity, local diffusion-barrier variations across the five-metal sublattice, and localized Ta partitioning, particularly at lower sintering temperatures. Possible C-site sublattice nonstoichiometry may also influence boundary migration; however, carbon vacancies and oxygen incorporation should not be treated as equivalent defect species. In carbide and oxycarbide ceramics, carbon vacancies and/or oxygen-related nonstoichiometry can modify diffusion pathways and may enhance mass transport and grain growth [23,24,33-35]. Therefore, the present work does not consider carbon vacancies to have a defect-pinning role. Because the present work did not independently quantify carbon and oxygen contents, the individual roles of carbon vacancies and oxygen incorporation cannot be separated from the present dataset. Accordingly, the apparent growth exponent and activation energy reported here are interpreted as effective kinetic parameters for this nominally carbon-deficient high-entropy oxycarbide composition, rather than as direct evidence for a specific carbon-vacancy- or oxygen-controlled grain-growth mechanism.

The apparent activation energy can be interpreted in terms of an effective grain-boundary mobility during the dwell time at the final hold temperatures in the SPS cycle. In normal grain growth, the growth factor is related to the product of grain-boundary mobility and grain-boundary energy; therefore, the measured Arrhenius dependence reflects the temperature sensitivity of boundary migration under the present processing conditions. In HEC ceramics, this effective mobility is expected to differ from that of simpler monocarbides because boundary migration occurs through a chemically complex lattice containing five transition-metal species on the metal sublattice. Local variations in bonding, atomic size, and diffusion barriers can reduce or broaden the distribution of boundary mobilities compared with a chemically simpler carbide. In addition, residual local partitioning, especially Ta-rich and Ta-deficient regions observed at lower SPS temperatures, may produce solute-drag-like effects that retard boundary motion. Possible C-site sublattice nonstoichiometry, including carbon vacancies and/or oxygen incorporation, may also affect boundary mobility by modifying local bonding and diffusion pathways. However, because the present work does not independently quantify carbon and oxygen contents, these effects cannot be separated and should be regarded as possible contributors rather than individually proven mechanisms. Thus, the reported activation energy represents an apparent mobility barrier for grain growth in this specific nominally carbon-deficient high-entropy oxycarbide system under SPS conditions.

A limitation of the present study is that direct carbon and oxygen measurements were not performed on the reacted powders or sintered ceramics. Therefore, the final values of x and y in $(Cr,Mo,Ta,V,W)C_xO_y$, as well as the corresponding carbon-vacancy concentration, cannot be independently determined from the present dataset. As a result, the discussion of carbon vacancies and oxygen incorporation is limited to their possible influence on grain-boundary mobility and should not be interpreted as a quantitative assignment of the defect chemistry.

The magnitude of the apparent activation energy is consistent with diffusion-controlled processes for refractory carbides. This interpretation is consistent with prior carbide and oxycarbide studies showing that C-site sublattice nonstoichiometry can affect densification and grain-growth kinetics, often by modifying diffusion-assisted mass transport rather than by simple defect pinning [23,24,26,27]. For instance, ZrC densification during SPS shows power-law creep-type behavior with apparent activation energies of ~563-576 kJ $mol^{-1}$ (vs ~653 kJ $mol^{-1}$ in hot pressing), while $ZrC_xO_y$ SPS densification is diffusion-assisted with high activation energies of ~687-774 $kJ \cdot mol^{-1}$ and is further accelerated by dissolved oxygen content and/or vacancy stoichiometry. The apparent activation energy from the present study provides a useful baseline kinetic parameter for multi-principal element carbides. Importantly, the densification results discussed above indicated that most samples reached full density before the final dwell, which helped decouple densification from subsequent coarsening and enables the temperature-dependent grain growth and homogenization trends to be interpreted without porosity as a confounding (dominant) factor.

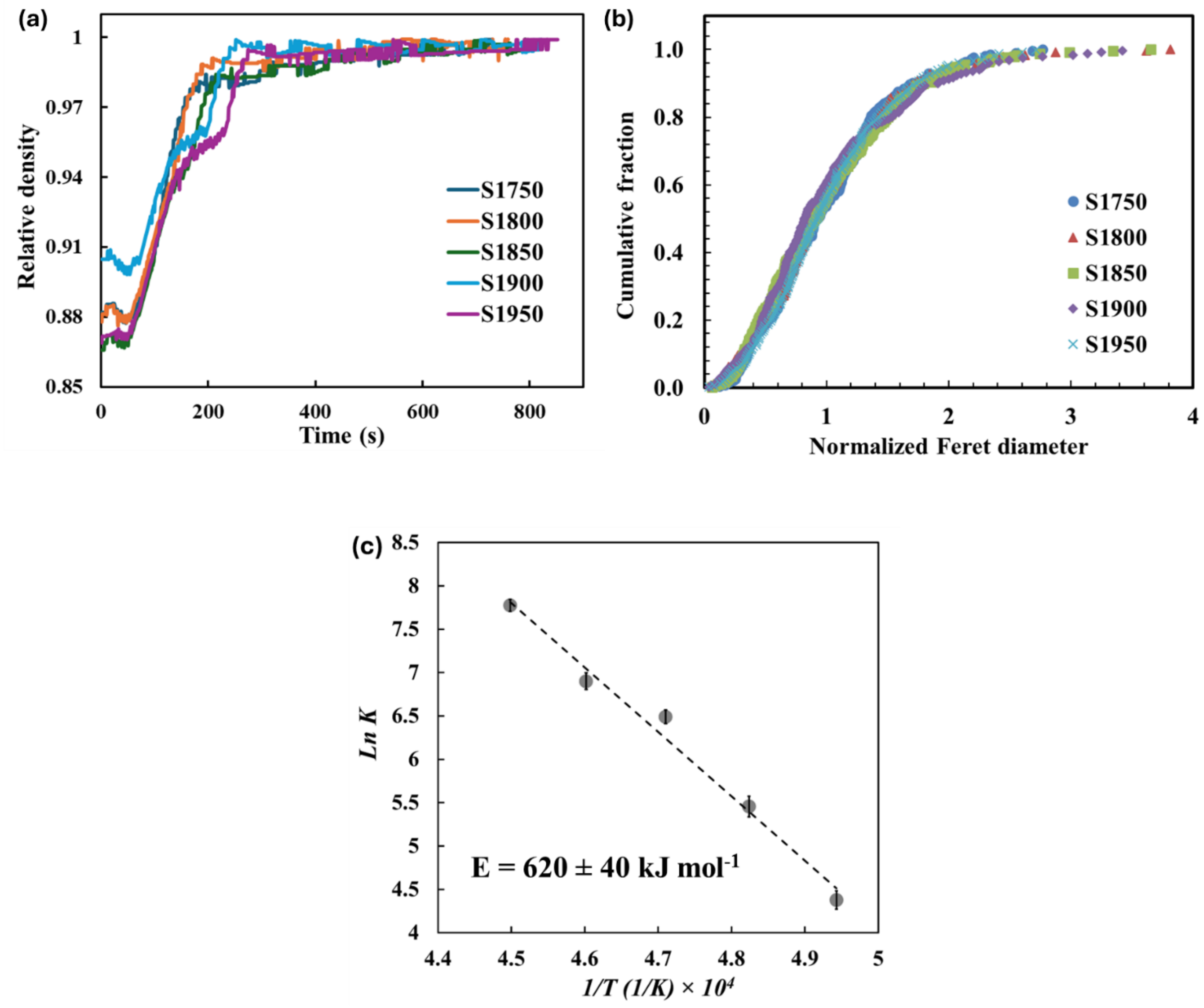

**Figure 5.** SPS sintering/densification behavior and grain growth kinetics. (a) SPS sintering/densification curves calculated from ram displacement data after thermal-expansion correction of the die assembly, showing the evolution of relative density during heating and dwell for the different final sintering temperatures. (b) Normalized cumulative grain-size distributions for (Cr,Mo,Ta,V,W)C specimens sintered at 1750-1950 °C. Feret diameters were normalized by the mean grain size of each specimen to assess approximate self-similarity of the measured distributions. (c) Arrhenius analysis of grain growth kinetics based on the growth factor $K$, plotted as $ln\,K$ versus $1/T$, with linear fitting used to estimate the apparent activation energy for grain-boundary migration.

## Conclusions

Near fully dense, single-phase (Cr,Mo,Ta,V,W)C high-entropy carbide ceramics were produced by carbothermal reduction followed by SPS at 1750-1950 °C using a constant dwell time of 10 min. Because all specimens achieved near-full density and retained the rock salt structure, the temperature series enabled direct assessment of SPS-temperature-driven microstructural evolution without the influence of porosity or phase changes. Increasing SPS temperature produced grain coarsening and a small but systematic increase in lattice parameter, together with improved chemical homogenization evidenced by reduced Ta segregation at higher temperatures. Grain growth analysis using a normal growth Arrhenius framework (with an assumed exponent of $n$=3) yielded an apparent activation energy of ~620 kJ $mol^{-1}$, consistent with diffusion-controlled grain-boundary migration in refractory carbides under SPS conditions. Densification curves indicated that most shrinkage occurred before the final dwell, while the dwell primarily promoted coarsening and equilibration, reinforcing SPS temperature as a key variable controlling grain-boundary mobility and diffusion-assisted homogenization in this HEC system. These results provide a quantitative kinetic basis for tailoring grain size and chemical uniformity in (Cr,Mo,Ta,V,W)C through SPS processing and motivate future time-dependent studies to independently determine the grain growth exponent and identify the rate-limiting diffusion pathway.

**Acknowledgements**

The authors acknowledge Missouri University of Science and Technology for providing the facilities and resources used in this work. The authors are grateful to Dr. Eric Bohannan at Missouri S&T for the helpful support and discussion with XRD. The authors thank Dr. Xiomara

Campilongo and Dr. Scott Thiel from Duke University for fruitful discussion. This research was supported by the Office of Naval Research under Award No. N00014-24-1-2768.

## Contributions

**A. Sarikhani** (Conceptualization, Methodology, Visualization, Validation, Writing - original draft, Data curation), **G. E. Hilmas** (Conceptualization, Supervision, Validation, Writing - review & editing), **D. W. Lipke** (Conceptualization, Validation, Writing - review & editing), **D. E. Wolfe** (Conceptualization, Writing - review & editing), **S. Curtarolo** (Conceptualization, Writing - review & editing), **S. J. Dillon** (Conceptualization, Writing - review & editing), **A. Mirzaei** (Conceptualization, Writing - review & editing), **W. G. Fahrenholtz** (Conceptualization, Methodology, Supervision, Validation, Writing - review & editing).

# *Supplementary Material*

## Grain Growth Kinetics in (Cr,Mo,Ta,V,W)C High-Entropy Carbide Ceramics

Ali Sarikhani[1,*], Gregory E. Hilmas[2], David W. Lipke[2], Douglas E. Wolfe[3], Stefano Curtarolo[4], Shen J. Dillon[5], Ahmad Mirzaei[5], William G. Fahrenholtz[2,*]

*[1] Missouri University of Science and Technology, Materials Research Center, Rolla, MO, USA*
*[2] Missouri University of Science and Technology, Materials Science and Engineering, Rolla, MO, USA*
*[3] Pennsylvania State University, Materials Science and Engineering, University Park, PA, USA*
*[4] Duke University, Mechanical Engineering and Materials Science, Center for Extreme Materials, Durham, NC, USA*
*[5] University of California, Irvine, Materials Science and Engineering, Irvine, CA, USA*

***ORCID:***

***Sarikhani**: 0000-0001-8121-0867*
***Hilmas**: 0000-0002-1611-5457*
***Lipke**: 0000-0002-4557-6690*
***Wolfe**: 0009-0004-5090-0290*
***Curtarolo**: 0000-0003-0570-8238*
***Dillon**: 0000-0002-6192-4026*
***Mirzaei**: 0000-0003-1518-9937*
***Fahrenholtz**: 0000-0002-8497-0092*

*S1. Microstructure and composition*

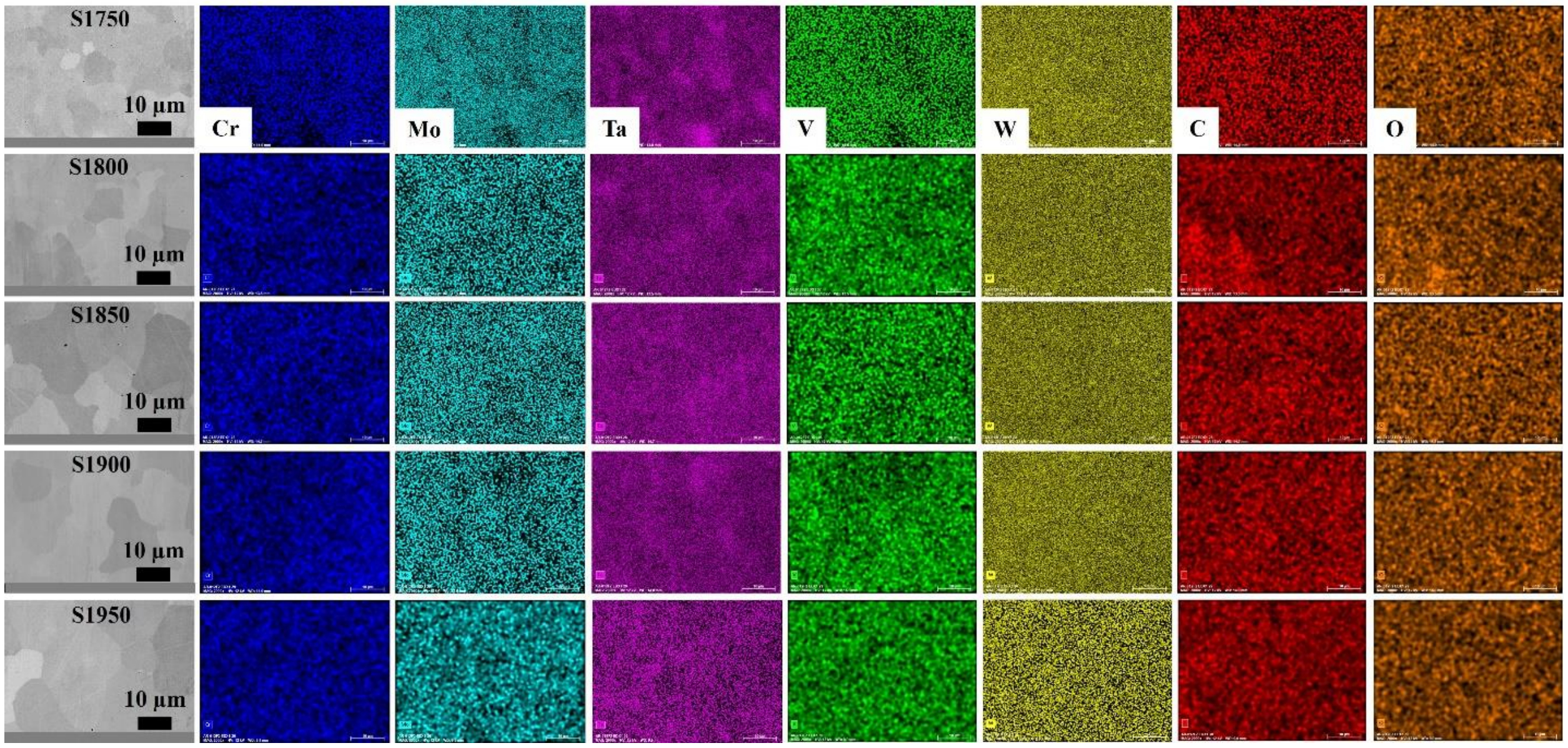

**Figure S1.** Same-magnification BSE images and EDS elemental maps for nominally carbon-deficient (Cr,Mo,Ta,V,W)C specimens sintered at 1750, 1800, 1850, 1900, and 1950 °C. Elemental maps are shown for Cr, Mo, Ta, V, W, C, and O using the same scale to enable direct comparison of spatial elemental distributions across the SPS temperature series. The maps provide qualitative spatial information on elemental uniformity, while quantitative EDS compositions and Ta-rich/Ta-deficient analyses are reported in Figs. 3 and 4.